# Toward CT-Equivalent Image Quality in Low-Dose Radiotherapy Planning: Conditional Diffusion–Based CBCT-to-CT Synthesis and the Impact of CBCT Input Representation

**Alzahra Altalib[1,2], Chunhui Li[2], Christopher Hamill Taylor[3], Sankar Pillai[2,3], Alessandro Perelli[4]**

[1] Faculty of Applied Medical Sciences, Jordan University of Science and Technology, Irbid, Jordan

[2] School of Science and Engineering, University of Dundee, Scotland, UK

[3] NHS Ninewells Dundee, Scotland, UK

[4] School of Cardiovascular and Metabolic Health, College of Medicine, Veterinary and Life Sciences, University of Glasgow, Scotland, UK

## 1. Purpose

During standard radiotherapy planning, repeated CT acquisitions are often required for patient registration, verification, and adaptive planning, resulting in increased cumulative X-ray dose. To mitigate this, low-dose cone-beam CT (CBCT) is routinely acquired during treatment delivery. However, CBCT image quality remains insufficient for accurate dose calculation and adaptive radiotherapy planning due to increased scatter, noise, beam hardening, and reconstruction-related artifacts [1, 2, 3].

This study develops a **supervised deep learning–based CBCT-to-CT synthesis framework** using a **conditional denoising diffusion probabilistic model (DDPM)** as in Fig. 1, where the generation of a CT-based planning for accurate positioning and dose calculation is obtained using generative models with low dose CBCT imaging.

Beyond demonstrating CBCT-to-CT synthesis, the primary objective is to investigate how the **representation of CBCT input data,** either standard clinical DICOM CBCT images or filtered back-projection (FDK) reconstructions from raw projection data, affects the performance of diffusion-based CT synthesis [2, 4].

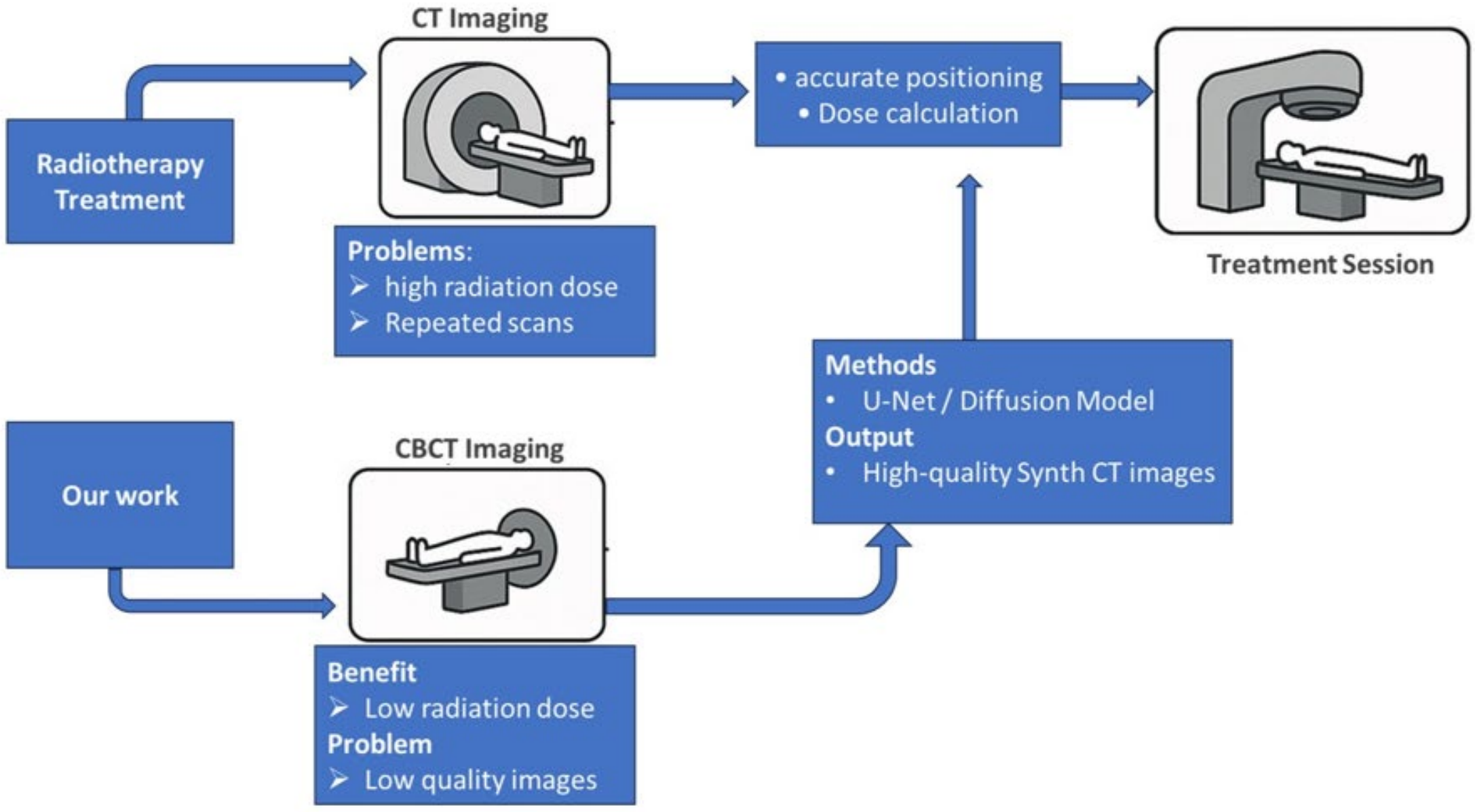

**Figure 1**: Proposed CBCT to CT synthesis workflow for radiotherapy. Low dose CBCT acquired during treatment is converted into high-quality synthetic CT using deep learning (U-Net / diffusion). This enables accurate positioning and dose calculation while reducing radiation exposure from repeated planning CT scans.

The overarching aim is to assess whether **physics-aware CBCT representations** better support CT-equivalent image quality while maintaining reduced imaging dose in radiotherapy workflows.

## 2. Methods and materials

### Background

CBCT image quality is influenced by scatter, beam hardening, noise, and reconstruction methodology. Standard clinical CBCT images stored as DICOM files are reconstructed using vendor-specific pipelines optimized for visualization rather than quantitative accuracy. In contrast, FDK is a well-established analytical reconstruction method that preserves the physical relationship between projection data and reconstructed voxel values, but with increased noise sensitivity [5].

Most deep learning methods for CBCT-to-CT synthesis use vendor-reconstructed DICOM CBCT as the model input. These CBCT volumes are primarily designed for clinical viewing, so they often include proprietary post-processing such as smoothing, scatter correction, and intensity normalization. While these steps can improve image appearance, they can also weaken or distort the physically meaningful link between the original projection measurements and the final voxel intensities. As a result, learning-based synthesis models may be limited by reconstruction decisions that are not intended to support quantitative CT equivalence, potentially constraining the quality and reliability of the conditioning information provided to the network.

Filtered back-projection using the Feldkamp-Davis-Kress (FDK) algorithm offers a contrasting reconstruction pathway. FDK works directly from raw CBCT projection data and compared with vendor-reconstructed DICOM images, typically applies minimal post-processing. Although FDK CBCT images can be noisier, they retain a clearer physical relationship between acquisition geometry, measured projections, and reconstructed voxel intensities. From the perspective of conditional diffusion, such reconstructions may provide more informative guidance signals for iterative denoising and may better support the suppression of acquisition-related artifacts an outcome diffusion models are explicitly suited to achieve.

**Method**

Diffusion-based generative models have recently gained attention for medical image synthesis because their iterative denoising formulation can produce high-quality outputs and often more stable training than many alternative generative frameworks. In conditional denoising diffusion probabilistic models (DDPMs), the network learns to iteratively denoise images while being guided by CBCT input data as shown in Fig. 2.

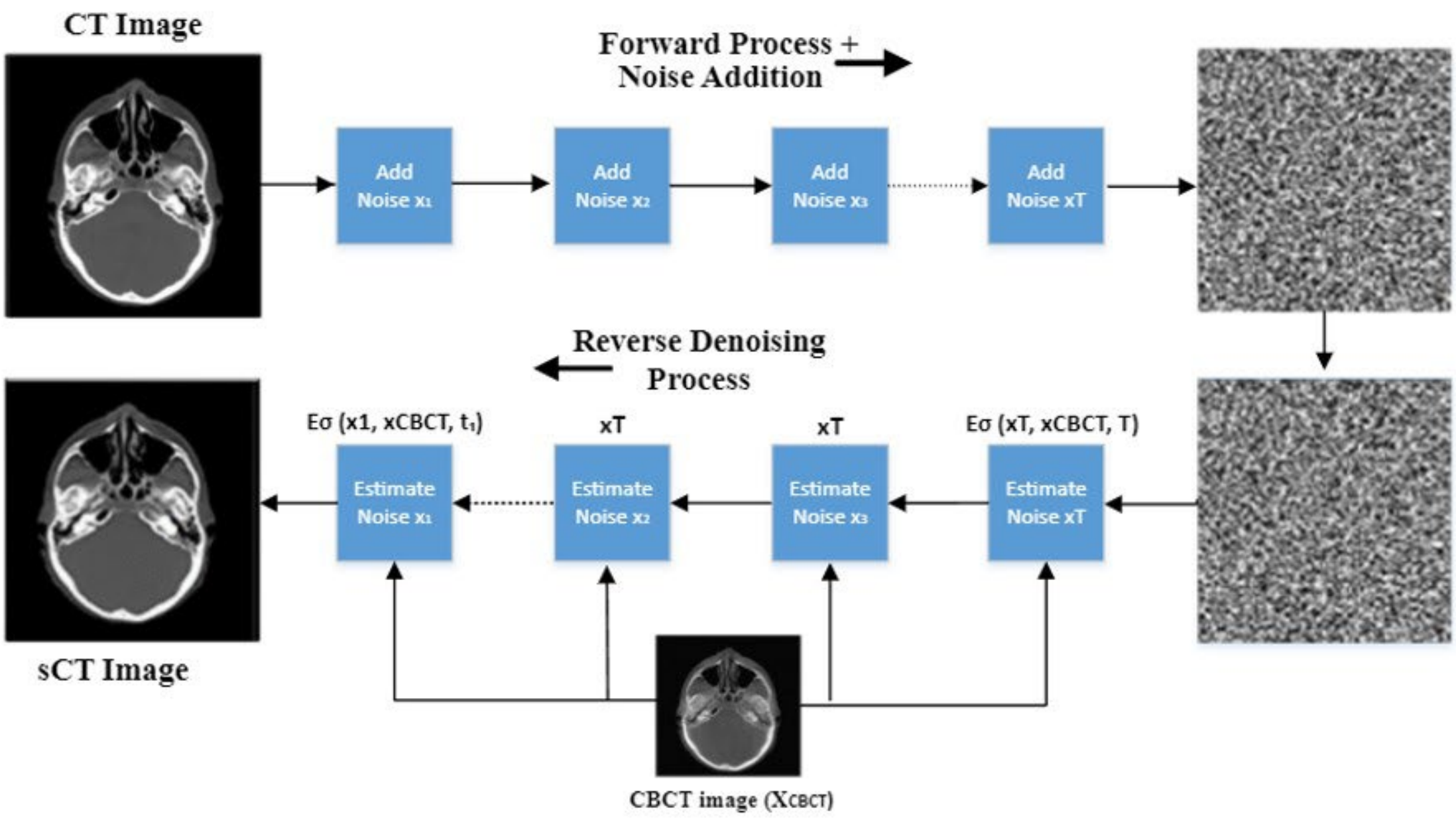

**Figure 2:** Conditional Diffusion Probabilistic Model (DDPM) for CBCT-to-CT image synthesis. During training, the model employs a diffusion-based framework with a forward process where noise is added to the input CT image and the reverse process where the sCT image is generated with CBCT guidance. In testing, only the reverse process is used with input random noise and CBCT guidance without the need of the CT image.

Consequently, both the statistical and physical properties of the conditioning input are critical, as they influence the denoising trajectory and the extent to which the resulting synthetic CT preserves anatomical detail and accurate intensity patterns. Despite this, most diffusion-based CBCT-to-CT studies continue to rely on standard vendor-reconstructed DICOM CBCT inputs and rarely examine whether the CBCT reconstruction methodology itself affects diffusion dynamics, artifact suppression, or overall synthesis fidelity [4, 6].

Diffusion models iteratively denoise images by reversing a predefined forward noise process and are therefore sensitive to the statistical and structural properties of the conditioning input. We hypothesize that CBCT images reconstructed using FDK from raw projection data provide more physically meaningful conditioning for diffusion-based CBCT-to-CT synthesis than standard DICOM CBCT.

The novelty of this work is in the systematic integration and evaluation of FDK-reconstructed CBCT as conditioning input for diffusion-based CBCT-to-CT synthesis in a radiotherapy setting. **To the best of our knowledge, this is the first study to quantitatively assess how CBCT reconstruction from raw projection data influences the performance of conditional diffusion models for synthetic CT generation.** By training and testing identical diffusion architectures using either DICOM CBCT or FDK-reconstructed CBCT as guidance, this study isolates the impact of CBCT input representation on synthetic CT fidelity under controlled conditions.

**Dataset and Imaging Setup**

A solid anthropomorphic head-and-neck phantom was scanned at Ninewells Hospital using matched CT and CBCT acquisition protocols. Ten repeated phantom scans yielded **920 aligned CT–CBCT slice pairs**, which were split into **80% training** (n = 736 slice pairs) and **20% independent testing** (n = 184 slice pairs), ensuring no overlap between sets. As input for the conditional DDPM, all CT images were normalized to a fixed Hounsfield Unit window (-1000 to 2000 HU) and resized to 256 x 256 pixels.

**Conditional Diffusion Model and Experimental Design**

A conditional DDPM with a U-Net backbone and multi-scale self-attention blocks was implemented. **Training** was performed for **1500 epochs** using a combined mean squared error (MSE) and structural similarity (SSIM) loss [4, 6,7]. Two experiments were conducted with **identical architecture, optimization strategy, and dataset split**, differing only in the CBCT representation used as conditioning during denoising:

1. Standard clinical CBCT provided as DICOM images.
2. CBCT reconstructed using FDK from raw projection data.

## 3. Results

**Training Phase Evaluation (Epoch 1500)**

Both configurations demonstrated stable convergence of the self-attention conditional DDPM. At the best-performing epoch, average training loss was comparable for DICOM-based input (0.00698) and FDK-based input (0.00664), indicating consistent optimization.

Direct comparison between original CBCT and ground-truth CT during training showed that:

- **DICOM CBCT** was initially closer to CT (**SSIM ≈ 0.83; PSNR ≈ 25.9 dB**), reflecting vendor-specific smoothing and artifact suppression [3, 5].
- **FDK CBCT** exhibited lower baseline similarity (**SSIM ≈ 0.26; PSNR ≈ 14.6 dB**) but reduced low-frequency bias and preserved physically consistent intensity relationships.

After diffusion-based synthesis, both approaches achieved substantial improvement relative to their CBCT baselines as in Fig. 3.

- **Synthetic CT** generated from **DICOM CBCT** reached **SSIM ≈ 0.93** and **PSNR ≈ 36.0 dB.**
- **Synthetic CT** generated from **FDK CBCT** achieved **SSIM ≈ 0.85** and **PSNR ≈ 37.4 dB.**

| Table ( A ) | Train Epoch 1500 Evaluation Metrics | |
|---|---|---|
| | DDPM using CBCT DICOM image | DDPM using CBCT FDK reconstructed |
| Number of images | 736 | 736 |
| Average Loss | 0.006981 | 0.006639 |
| Subset Evaluation | | |
| CT vs Synth CT | | |
| →AVG SSIM | 0.9315 | 0.8458 |
| →AVG PSNR (dB) | 36.00 | 37.41 |
| →AVG MSE | 0.001414 | 0.000684 |
| →AVG MAE | 0.010500 | 0.008770 |
| CBCT vs CT | | |
| →AVG SSIM | 0.8328 | 0.2605 |
| →AVG PSNR (dB) | 25.89 | 14.61 |
| →AVG MSE | 0.003712 | 0.034860 |
| →AVG MAE | 0.024696 | 0.169841 |
| Improvement from training for CBCT to Synth CT | | |
| →AVG SSIM | 0.0987 | 0.5853 |
| →AVG PSNR (dB) | 10.11 | 22.8 dB |

| Table ( B ) | Result (Testing) Used checkpoint 1500 | |
|---|---|---|
| | DDPM using CBCT DICOM image | DDPM using CBCT FDK reconstructed |
| Number of images | 184 | 184 |
| Test metrics/_mean_std | | |
| CT vs Synth CT | | |
| →AVG SSIM+ std+ std | 0.9465 ± 0.0659 | 0.8168 ± 0.1653 |
| →AVG PSNR (dB)+ std | 37.35 dB ± 7.14 | 34.49 dB ± 6.51 |
| →AVG MSE+ std | 0.000885 ± 0.001613 | 0.000927 ± 0.001048 |
| →AVG MAE+ std | 0.008452 ± 0.007335 | 0.010737 ± 0.006879 |
| CBCT vs CT | | |
| →AVG SSIM+ std+ std | 0.8302 ± 0.0759 | 0.2628 ± 0.0710 |
| →AVG PSNR (dB)+ std | 25.57 dB ± 3.55 | 14.41 dB ± 0.35 |
| →AVG MSE+ std | 0.003683 ± 0.002439 | 0.036371 ± 0.003110 |
| →AVG MAE+ std | 0.024596 ± 0.009695 | 0.173848 ± 0.010854 |
| Improvement from testing for CBCT to Synth CT | | |
| →AVG SSIM | 0.1163 | 0.554 |
| →AVG PSNR (dB) | 11.78 | 20.08 dB |

**Figure 3:** Quantitative performance comparison for DDPM training (Table A) and testing (Table B). Higher median SSIM and PSNR are observed when FDK reconstructed raw CBCT data is used as guidance.

These results confirm that the diffusion model effectively suppresses noise and artifacts for both CBCT representations during training, while preserving anatomical structure as in Fig. 4.

**Testing Phase Evaluation**

Independent **testing on 184** previously unseen CT–CBCT pairs confirmed the training trends. **Baseline DICOM CBCT showed higher similarity to CT than FDK CBCT**; however, diffusion-based synthesis markedly improved image quality in both cases.

**FDK-conditioned synthetic CT** achieved **SSIM = 0.817 ± 0.165** and **PSNR = 34.49 ± 6.51 dB** as shown in Fig. 3. Relative to original CBCT input, this corresponds to an average improvement of **+0.55 SSIM** and **+20.1 dB PSNR**, compared with **+0.12 SSIM** and **+11.8 dB PSNR** when conditioning on **DICOM CBCT** as in Figs. 5-6.

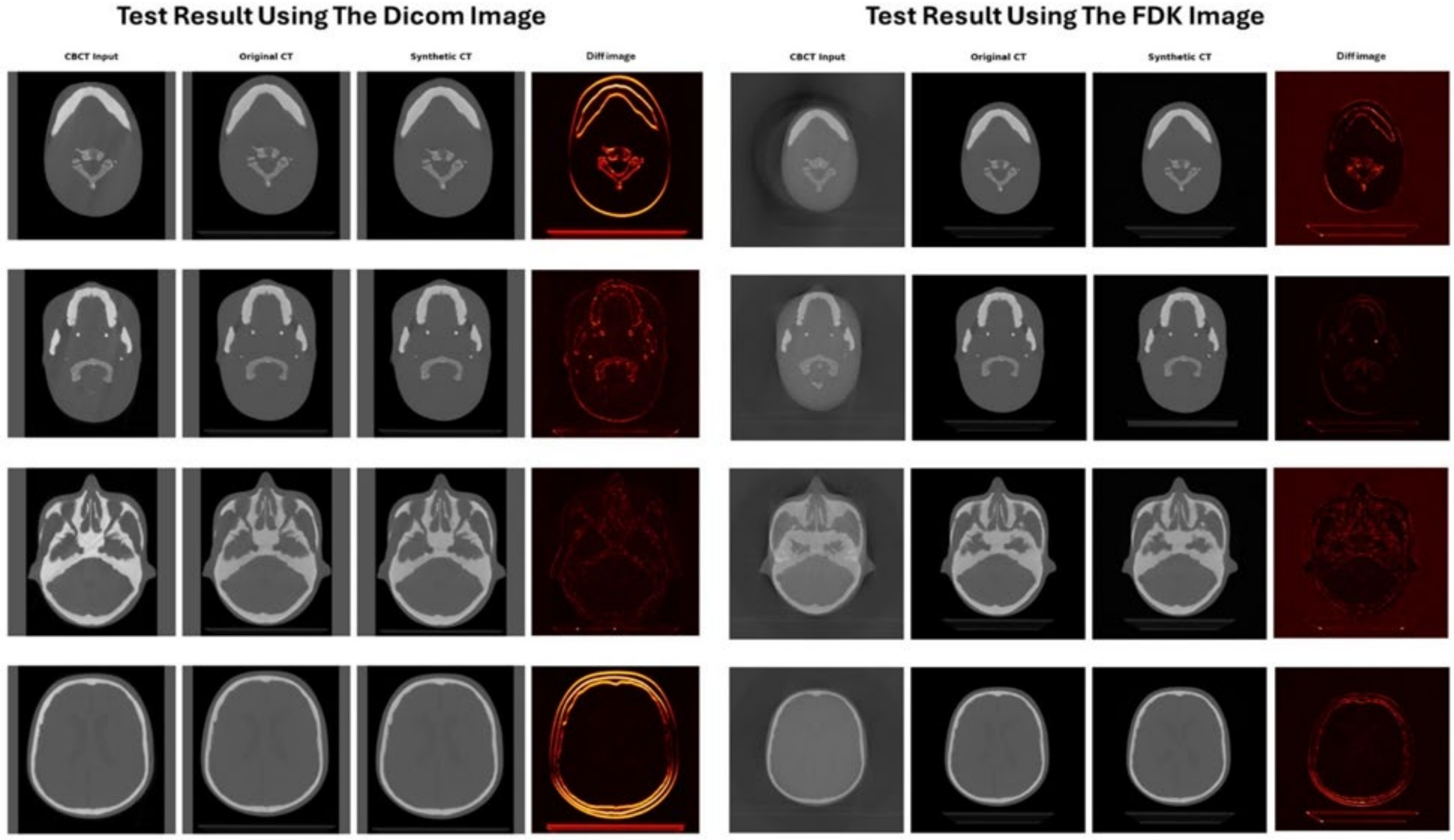

**Figure 4:** visual comparison of CBCT input, ground truth CT, synthetic CT, and error difference using DDPM using (A) CBCT DICOM image guidance, raw CBCT data. The FDK conditioned output (B) shows improved structural consistency.

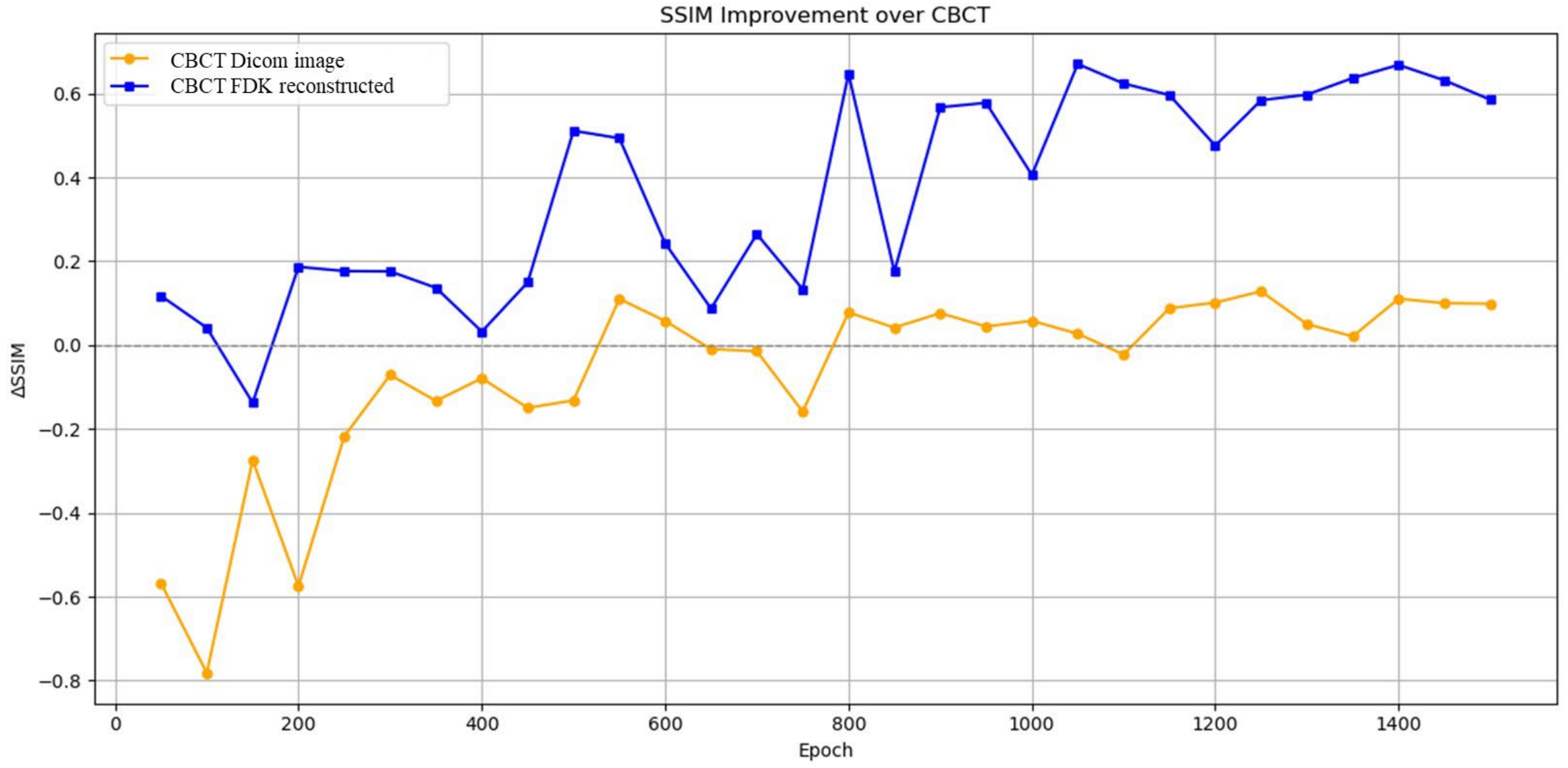

**Figure 5:** ΔSSIM over baseline CBCT during training. The figure shows the structural similarity improvement of synthesized CT over original CBCT. Conditioning on FDK reconstructions from raw data yields consistently higher ΔSSIM than CBCT DICOM input, indicating better structural fidelity in the generated CT.

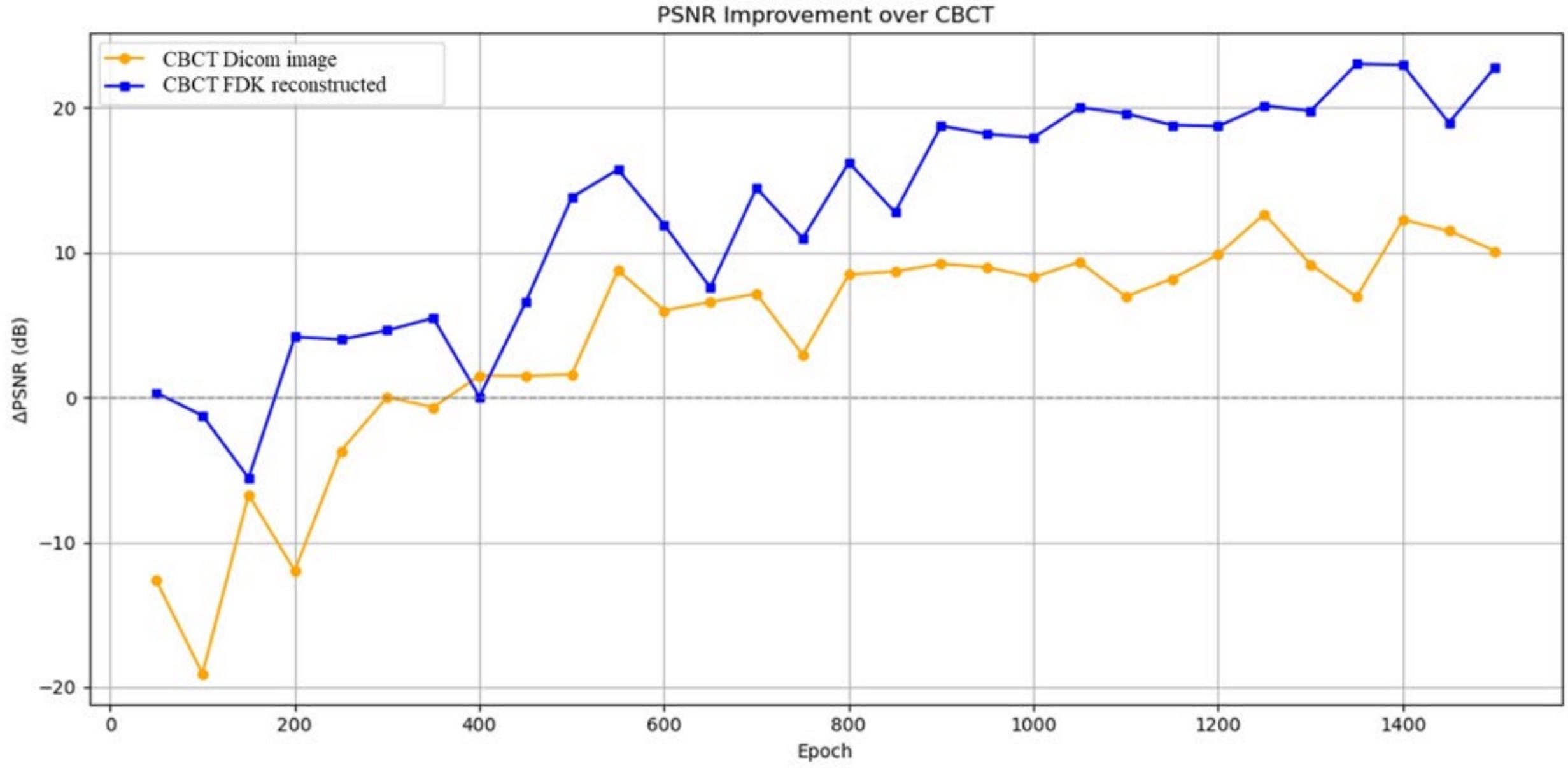

**Figure 6:** ΔPSNR over baseline CBCT during training. The figure shows the peak signal to noise ratio improvement of synthesized CT relative to original CBCT. Conditioning on FDK reconstructions from raw projection data consistently achieves higher ΔPSNR than CBCT DICOM input, indicating reduced noise and improved fidelity in the synthesized CT.

**Interpretation and Key Finding**

These results demonstrate that:

- **DICOM CBCT appears closer to CT before denoising**, due to smoothing and vendor post-processing.
- **FDK CBCT, although noisier, provides more physically meaningful conditioning** for the diffusion model.
- As a result, **diffusion-based denoising yields greater improvement when conditioned on FDK CBCT**, highlighting the importance of reconstruction-aware CBCT representations.

This finding explains why FDK-conditioned diffusion achieves stronger relative enhancement and supports the integration of raw-data based CBCT reconstructions in advanced CBCT-to-CT synthesis pipelines [7,8].

**Qualitative Evaluation**

Visual inspection demonstrated that synthetic CT images generated using **FDK-conditioned CBCT** exhibited:

- Reduced streak and cupping artifacts.
- Improved soft-tissue uniformity.
- Enhanced structural consistency at bone-air interfaces.

Difference maps confirmed lower residual error for the **FDK-conditioned synthesis** compared with the **DICOM-conditioned** configuration.

## 4. Conclusion

The representation of CBCT images used during conditional diffusion significantly influences CBCT-to-CT synthesis performance. While standard DICOM CBCT appears closer to CT at baseline due to vendor post-processing, FDK-reconstructed CBCT from raw projection data provides more physically meaningful conditioning for diffusion models. As a result, conditional diffusion-based synthesis achieves substantially greater relative improvement when conditioned on FDK CBCT.

**These findings highlight the importance of reconstruction aware AI pipelines and support the integration of raw-data-based CBCT representations to enhance synthetic CT generation for radiotherapy imaging [4, 7, 8].** However, access to clinical raw CBCT projection datasets remains limited. Future work will focus on patient-based validation, multi-scanner evaluation, and assessment of dosimetric impact in adaptive radiotherapy workflows.

## Personal information

Alzahra Altalib is a third-year PhD researcher at the University of Dundee, UK, specializing in medical physics for radiation therapy, with a particular focus on CT image synthesis. Her research explores advanced image generation techniques to enhance treatment planning and diagnostic accuracy in radiotherapy.

Alzahra Altalib holds an MSc in Medical Physics (2023) from the University of Aberdeen and a BSc in Radiologic Technology (2021) from Jordan University of Science and Technology.

With over 22 years of professional experience in the Royal Medical Services, Alzahra Altalib has held several leadership roles, including CT Supervisor at King Hussein Medical Center and Head of Radiography at Queen Alia Military Hospital in Amman.

## References

[1] Kurz, C., Kamp, F., Park, Y.K., Zöllner, C., Rit, S., Hansen, D., Podesta, M., Sharp, G.C., Li, M., Reiner, M. and Hofmaier, J., 2016. Investigating deformable image registration and scatter correction for CBCT-based dose calculation in adaptive IMPT. *Medical physics*, *43*(10), pp.5635-5646.

[2] Spadea, M.F., Maspero, M., Zaffino, P. and Seco, J., 2021. Deep learning based synthetic-CT generation in radiotherapy and PET: a review. *Medical physics*, *48*(11), pp.6537-6566.

[3] Çağlar, M., Ertaş, K.S., Cebe, M.S., Kara, I., Kheradmand, N. and Metcalfe, E., 2025. Synthetic CT generation from CBCT using deep learning for adaptive radiotherapy in prostate cancer. *Frontiers in Radiology*, *5*, p.1680803.

[4] Altalib, A., Li, C. and Perelli, A., 2026. Conditional diffusion models for CT image synthesis from CBCT: A systematic review. Tomography, 12(5), p.64.

[5] Buzug, T.M., 2009. Computed tomography: from photon statistics to modern cone-beam CT. *Soc Nuclear Med*, *50*(7).

[6] Ho, J., Jain, A. and Abbeel, P., 2020. Denoising diffusion probabilistic models. *Advances in neural information processing systems*, *33*, pp.6840-6851.

[7] Altalib, A., Li, C. and Perelli, A., 2026. Equivariant conditional diffusion model for head and neck CT image synthesis from CBCT. Medical Physics, 53(8), p.e70617.

[8] Altalib, A., McGregor, S., Li, C. and Perelli, A., 2025. Synthetic CT image generation from CBCT: a systematic review. IEEE Transactions on Radiation and Plasma Medical Sciences, 9(6), pp.691-707.